\documentclass{llncs}
\usepackage{llncsdoc}

\usepackage[usenames,dvipsnames]{color}
\usepackage{framed}
\usepackage{comment}
\usepackage{lipsum}
\usepackage{ntheorem}

\newenvironment{counterex}[1][Counterexample]{\begin{trivlist}
\item[\hskip \labelsep {\bfseries #1}]}{\end{trivlist}}

\includecomment{Long}
\includecomment{Extended}
\excludecomment{Short}
\usepackage[toc,page]{appendix}

\usepackage{graphicx}
\usepackage{cite}
\usepackage{url}
\usepackage{amsmath}
\usepackage{amssymb}
\usepackage{pgf}
\usepackage{pgfplots}
\usepackage{tikz}
\usetikzlibrary{arrows,automata}
\definecolor{light-gray}{gray}{0.8}
\newtheorem*{Openquest}{Open question}

\usepackage{microtype}

\title{
A Note on a Recent Attempt to Improve the Pin-Frankl Bound
}

\author{Fran\c cois Gonze \inst{1} \and Rapha\"el M. Jungers \inst{1} \thanks{R. M. Jungers is a F.R.S.-FNRS Research Associate}
\thanks{This work was also supported by the communaut\'e francaise de Belgique - Actions de Recherche Concert\'ees and by the Belgian Program on Interuniversity Attraction Poles initiated by the Belgian Federal Science Policy Office.} \and A.N. Trahtman \inst{2}}

\institute{ICTEAM Institute\\
UCLouvain, Louvain La Neuve, Belgium\\
 \email{$\{$francois.gonze,raphael.jungers$\}$@uclouvain.be}\\
\and 
 Dep. of Math.\\
Bar-Ilan University, 52900, Ramat Gan, Israel\\
 \email{trakht@macs.biu.ac.il} 
 }

\begin{document}
\maketitle
\begin{abstract}

We provide a counterexample to a lemma used in a recent tentative improvement of the the Pin-Frankl bound for synchronizing automata. This example naturally leads us to formulate an open question, whose answer could fix the line of proof, and improve the bound.

\end{abstract}

\section*{A Counterexample}

This short note studies a problem related with synchronizing automata and \v Cern{\'y}'s conjecture, formulated in \cite{Cerny64}.
A good survey on the topic is given in \cite{volkov_survey}. See \cite{JungersBlondelOlshevsky14}, \cite{ArxivPrep}, \cite{jungers_sync_12} for recent work on the subject.

A (deterministic, finite state, complete) \emph{automaton} (DFA) is a triplet $(Q,\Sigma, \delta)$ with $Q$ the set of \emph{states}, $\Sigma$ the alphabet of \emph{letters} and $\delta$ the transition function $\delta: Q\times \Sigma \rightarrow Q$ defining the effect of the letters on the states. For $q_i, q_j \in Q$ and $l \in \Sigma$, we write $q_il=q_j$ if $\delta(q_i, l)=q_j$. We call a \emph{word} $w$ of \emph{length} m a sequence of $m$ letters $l_1 ... l_m$, $l_i \in \Sigma, 1\leq i\leq m$. We write $\Sigma^m$ the set of words of length m. For $q_i, q_j \in Q$ and $w= l_1 ... l_m \in \Sigma^m$, we write $q_i w=q_j$ if $\delta(...\delta(\delta(q_i, l_1),l_2)...,l_m)=q_j$. For an automaton with $n$ states and a word $w$, we note $Qw=\{q_j|q_iw=q_j, 1\leq i \leq n\}$ the set of states that are in the image of $w$. We can represent an automaton as a directed graph. Each state is represented as a vertex, and the effect of each letter on each state is represented as a directed edge. We call a DFA \emph{strongly connected} if its graph representation is a strongly connected graph.

A word $w$ is called \emph{synchronizing word} if, for any states $q_i, q_j \in Q$, $q_iw=q_jw$. A DFA is called \emph{synchronizing automaton} if it has a synchronizing word.

\v Cern{\'y}'s conjecture \cite{Cerny64} states that any \emph{synchronizing automaton} with $n$ states has a \emph{synchronizing word} of length at most $(n-1)^2$.

So far the best proven bound is $(n^3-n)/6$, obtained more than 30 years ago in \cite{Frankl82} and \cite{Pin83a}, and re discovered independently in \cite{KlyachkoRystsovSpivak87}. Recently, a tentative improvement to $n(7n^2+6n-16)/48$ has been proposed in  \cite{Trahtman2011}. However, as mentioned later by the author on ArXiv \cite{ArxivPTraht}, there is a flaw in the proof. Nevertheless, since the publication of \cite{Trahtman2011}, many new papers are citing this result, and no publication clearly confirms that the proof is not valid. In this note, we make this point clear by providing a counterexample to Lemma 3 in \cite{Trahtman2011}. The lemma is the following:

\begin{lemma}[Lemma 3 in \cite{Trahtman2011}]
\label{FalseLemma}
Let $Q$ be the set of states of a synchronizing strongly connected $n-$state DFA. Then for any state $q$ there exists a word $w$ of length not greater than $n$ such that $q \notin Q w$. For any $k<n$ there are at least k states $q_1, ..., q_k$ and words $w_1, ..., w_k$ of length not greater than $k$ such that $q_i\notin Q w_i$, $1\leq i\leq k$.
\end{lemma}

We contradict the lemma by exhibiting an automaton such that, for one state $q_0$, there is no word $w$ of length smaller or equal to $n$ with the property that $q_0 \notin Q w$.

\begin{counterex}

The automaton represented in Fig.\ref{Counter} is a synchronizing automaton, as the word $abbababba$ is a synchronizing word. However, the shortest word $w$ such that $q_0 \notin Q w$ is $w=abbaba$. Since the automaton has only 4 states and $t$ is 6 letters long, this contradicts Lemma \ref{FalseLemma}.

\begin{figure}[h!]
\begin{center}
\scalebox{1}{
\begin{tikzpicture}[->,>=stealth',shorten >=1pt,auto,node distance=1.7cm,
                    semithick]
  \tikzstyle{every state}=[fill=light-gray,draw=none,text=black, scale=0.9]

  \node[state] (A)                    {$q_0$};
  \node[state]         (B) [above right of=A] {$q_1$};
  \node[state]         (D) [below right of=A] {$q_3$};
  \node[state]         (C) [below right of=B] {$q_2$};

  \path (A) edge              	node {a} (B)
            edge [loop left]  	node {b} (C)
        (B) edge   				node {} (A)
            edge              	node {b} (C)
        (C) edge              	node {b} (D)
         	edge [loop right] 	node {a} (D)
        (D) edge  			  	node {b} (B)
        	edge  			  	node {a} (A);
\end{tikzpicture}}
\end{center}
\caption{A counterexample to Lemma \ref{FalseLemma}}
\label{Counter}
\end{figure}
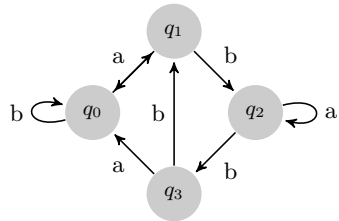

\end{counterex}

Lemma \ref{FalseLemma} was a key step in the improvement on the maximal length of a shortest synchronizing word. We observe that a weaker version of Lemma 1 could still improve the Pin-Frankl bound. In fact, any value proportional to the number of states of the automaton would lead to an improvement of the bound. This motivates us to raise the following open question.

\begin{Openquest}
Let $Q$ be the set of states of a synchronizing strongly connected $n-$state DFA.

Is there a constant $c$ such that, for any state $q\in Q$, there exists a word $w$ of length not greater than $cn$ such that $q \notin Q w$?
\end{Openquest}

\bibliographystyle{eptcs}
\bibliography{references}

\textit{•}

\end{document}